\documentclass{article}

     \PassOptionsToPackage{numbers, compress}{natbib}

     \usepackage[preprint]{neurips_2020}

\usepackage{graphicx} 
\usepackage{subfigure}
\usepackage[utf8]{inputenc} 
\usepackage[T1]{fontenc}    
\usepackage{hyperref}       
\usepackage{url}            
\usepackage{booktabs}       
\usepackage{amsfonts}       
\usepackage{nicefrac}       
\usepackage{microtype}      
\usepackage{enumitem}
\newlist{steps}{enumerate}{1}
\setlist[steps, 1]{label = step \arabic*:}

\usepackage{verbatim}
\usepackage[noend]{algpseudocode}
\usepackage{algorithmicx,algorithm}
\usepackage{amsmath}
\usepackage{amssymb}
\usepackage{threeparttable}

\title{Entropy Enhanced Multi-Agent Coordination Based on Hierarchical Graph Learning for Continuous Action Space}

\author{%
  Yining Chen, ~~Ke Wang, ~~Guanghua Song  \\
  School of Aeronatuics and Astronautics, Zhejiang University\\
  38 Zheda RD, Hangzhou, P.R. China\\
  \texttt{\{ch19930611, kwang\_0228, ghsong\}@zju.edu.cn} \\
  \And Xiaohong Jiang  \\
  College of Computer Science and Technology, Zhejiang University  \\
  38 Zheda RD, Hangzhou, P.R. China\\
  \texttt{jiangxh@zju.edu.cn} \\
}

\begin{document}

\maketitle

\begin{abstract}
  In most existing studies on large-scale multi-agent coordination, the control methods aim to learn discrete policies for agents with finite choices. They rarely consider selecting actions directly from continuous action spaces to provide more accurate control, which makes them unsuitable for more complex tasks. To solve the control issue due to large-scale multi-agent systems with continuous action spaces, we propose a novel MARL coordination control method that derives stable continuous policies. By optimizing policies with maximum entropy learning, agents improve their exploration in execution and acquire an excellent performance after training. We also employ hierarchical graph attention networks (HGAT) and gated recurrent units (GRU) to improve the scalability and transferability of our method. The experiments show that our method  consistently outperforms all baselines in large-scale multi-agent cooperative reconnaissance tasks.
\end{abstract}

\section{Introduction}\label{sec1}

The multi-agent system has been applied in various domains during the last decade \cite{dorri2018multi}, such as compute games \cite{rashid2018qmix, arulkumaran2019alphastar}, smart grids \cite{nguyen2012agent, hu2022multi}, and the unmanned aerial vehicle (UAV) navigation \cite{liu2018energy, qie2019joint}. By achieving consensus among agents, the multi-agent system can coordinate them to solve complex tasks efficiently in the real world. However, most previous works in large-scale multi-agent systems simplify the control problem of agents by discretizing their actions as finite choices. Although such methods may help agents to learn a stable policy, it is still hard for discrete action spaces to model their actions accurately in some complex scenarios. To address the challenges in accurate control, we focus on large-scale multi-agent systems with continuous action space and employ multi-agent reinforcement learning (MARL) to coordinate agents for various scale tasks.

In standard MARL, agents learn their policies in two ways - off-policy and on-policy. The off-policy approaches \cite{lillicrap2015continuous, fujimoto2018addressing, barth2018distributed} utilize the transitions stored in replay buffers to optimize deterministic policies, which brings good sample efficiency. However, they cannot prevent agents from converging to non-optimal policies in continuous action spaces, because they use the maximization based on Q-learning \cite{watkins1992q} for training. As a result, the instability in these policies limits the applicability of off-policy learning in large-scale multi-agent systems. The on-policy approaches \cite{mnih2016asynchronous, schulman2015trust, heess2017emergence, schulman2017proximal} derive more stable and robust policies by outputting the distribution of actions. Since they control the exploration by the stochasticity of their policies, the agents obtain better performance when executing real-world tasks. Unfortunately, these approaches require new samples for updating policies, meaning they suffer from poor sample complexity. Moreover, this shortcoming becomes extremely severe as the scale of multi-agent systems increases since agents need more samples to learn an effective policy.

We herein propose an MARL approach named Entropy-enhanced Hierarchical graph Continuous Action Multi-Agent coordination control method (EHCAMA), which contains a graph learning network and an entropy-enhanced optimization for training. To derive efficient and stable policies in continuous action spaces, we implement EHCAMA on an actor–critic framework and train the stochastic actor with an off-policy training strategy. We introduce maximum entropy learning \cite{haarnoja2017reinforcement} to augment the MARL training process with entropy maximization, which provides intelligent exploration to agents and helps them to find the optimal policies. At the same time, we adopt hierarchical graph attention networks (HGATs) \cite{ryu2020multi} and gated recurrent units (GRUs) \cite{cho2014properties} to summarize the information from agents' observations, neighbors, and memories, improving its scalability and transferability in large-scale environments.

The rest of this paper is organized as follows. We discuss related work in Section \ref{sec2} and introduce the background about MARL, HGAT, and maximum entropy learning in Section \ref{sec3}. In Section \ref{sec4}, we present a detailed description of our method. The experimental results are shown in Section \ref{sec5}. Finally, we conclude the paper in Section \ref{sec6}.

\section{Related Work}\label{sec2}

Deep reinforcement learning (DRL) is an efficient paradigm to address the coordination issues in multi-agent learning. It combines reinforcement learning (RL), which models the interaction between agents and environments, and deep neural networks (DNNs) that process high-dimensional observation data. The value-based DRL algorithms like deep Q-network (DQN) \cite{mnih2015human} compute Q-value for each action and select the one with the maximum Q-value. Therefore, they are only applicable to discrete action spaces. The actor-critic algorithms use a separate actor network to approximate the policies and directly adjust its parameters to maximize the expected rewards, which are feasible in continuous action space. Deep deterministic policy gradient (DDPG) \cite{lillicrap2015continuous}, a popular actor-critic method, employs off-policy learning to derive deterministic policies for agents. However, it is hard for DDPG to resolve complex tasks since its policies are unstable and non-robust to hyper-parameter settings. On the contrary, the on-policy algorithms, such as trust region policy optimization (TRPO) \cite{schulman2015trust} and proximal policy optimization (PPO) \cite{heess2017emergence, schulman2017proximal}, train stochastic policies with newly collected samples, which brings stability to agents but reduces their sample efficiency. The authors of \cite{heess2015learning} studied off-policy stochastic control and presented a zero-step variant of stochastic value gradients (SVG(0)) that optimizes the policy with a experience database. In \cite{haarnoja2018soft2}, Haarnoja et al. introduced maximum entropy learning into policy gradient method and proposed an off-policy algorithm named soft actor-critic (SAC). By incorporating an entropy term in the maximum expected return objective, it provides tractable stochastic policies for agents and encourages them to explore.

Unlike the above RL methods that only consider a single agent, MARL focuses on learning decentralized policies and controls the multi-agent system by achieving consensus among agents. A widely used solution in MARL is optimizing each agent's policy individually on a framework of centralized training and decentralized execution (CTDE), such as multi-agent deep deterministic policy gradient (MADDPG) \cite{lowe2017multi} and counterfactual multi-agent (COMA) \cite{foerster2018counterfactual}. These methods derive a decentralized policy for each agent with the gradient from their centralized critics. Following SAC, multi-actor-attention-critic (MAAC) \cite{iqbal2019actor} adopts maximum entropy learning to control the exploration. Furthermore, it constructs an attention network in the critic to quantify the importance of each agent when learning a value function. However, the individual optimization methods are infeasible for large-scale multi-agent systems for two reasons. First, the input space of their networks grows dramatically as the scale of multi-agent systems increases, which leads to poor scalability in various scale environments. Second, they do not consider the transferability of trained policies, so they have to retrain agents before executing new tasks.

To resolve the two issues mentioned above, another type of MARL method applies DNNs with shared parameters to derive decentralized policies for a set of homogeneous agents, while imposing their consensus through communication \cite{foerster2016learning}. The authors in \cite{ryu2020multi} designed hierarchical graph attention networks (HGATs) to model the inter-agent and inter-group relationships in the environment and summarize each agent's status into an embedding vector by HGAT, learning deterministic policies in an MADDPG framework. Our earlier work \cite{ren2022space} presented a multi-UAV navigation approach for space-air-ground integrated mobile crowdsensing. We employed convolutional neural networks (CNNs) to process observations while sharing information through a graph attention networks (GATs) \cite{velivckovic2017graph} based communication. Moreover, we designed a maximum entropy MARL algorithm based on the soft deep recurrent graph network (SDRGN) \cite{ye2022multi}. In another previous work \cite{chen2022scalable}, we proposed a scalable and transferable MARL method for large-scale multi-agent systems in mixed cooperative-competitive environments. It adopts HGAT to extract features from observations while recording long-term historical information in its recurrent unit, which provides outstanding performance in large-scale cooperative and competitive tasks with discrete action spaces.

Our proposed EHCAMA differs from the previous works \cite{ren2022space} and \cite{chen2022scalable} in that: (1) \cite{chen2022scalable} employs $\epsilon$-greedy or $\epsilon$-categorical strategy to provide exploration in discrete action spaces for agents, whereas EHCAMA derives stochastic policies for continuous action spaces and optimizes them with maximum entropy learning; (2) \cite{ren2022space} outputs the distribution of actions according to Q-values and trains Q-network based on soft Q-learning \cite{haarnoja2017reinforcement}, while we implement EHCAMA on the SAC framework and compute actions by a separate actor network; (3) \cite{ren2022space} regards the observation as a pixel map. EHCAMA represents the global state as a graph and stores it as adjacency matrices in a replay buffer, which is more advantageous in terms of space complexity.

\section{Background}\label{sec3}

\subsection{Multi-agent Reinforcement Learning (MARL)}

The process of MARL can be described as decentralized partially observable Markov decision process (Dec-POMDP) \cite{bernstein2002complexity}, which is defined as follow: $s$ represents the global state of the environment; $o_i$ denotes agent $i$'s local observation; $a_i$ is the action of agent $i$. Each agent optimizes its policy $\pi_i$ to maximize the total expected return $R_i=\sum_{t=0}^{T}\gamma^t r_i(t)$, where $T$ is a final timeslot and $r_i(t)$ is the reward that agent $i$ obtains at timeslot $t$. The discount factor $\gamma \in [0,1]$.

DDPG \cite{lillicrap2015continuous} is an off-policy actor-critic RL algorithm. It consists of an actor network $\mu$, which outputs an action according to a deterministic policy, and a critic network $Q$, estimating the total expected return. $Q$ is trained by minimizing $\mathcal{L}(\theta^Q)= \mathbb{E}[(y-Q(s,a|\theta^Q))^2]$, where $y=r+ \gamma Q'(s',a'|\theta^{Q'})$. $\mu$ is optimized by the gradient $\nabla_{\theta^{\mu}} J(\theta^{\mu})=\mathbb{E}[\nabla_{\theta^{\mu}}\mu(s|\theta^{\mu})\nabla_a Q(s,a|\theta^Q)|_{a=\mu(s|\theta^{\mu})}]$.

\subsection{Hierarchical Graph Attention Network (HGAT)}

HGAT is a novel network structure in MARL, which aims to extract the relationships among different types of agents. The idea of HGAT is to represent the global state as a graph and process it through a network that stacks multiple GATs. After receiving observations from the environment, agent $i$ uses GAT to aggregate embedding vectors $e_j$ from its neighbors in each group $l$ into ${e'}_i^l=\sum_{j \in \mathcal{N}^l_i} \alpha_{ij} \textbf{W}^l_v e_j$, where $\mathcal{N}^l_i$ denotes the set of agent $i$'s neighbors in group $l$. The attention weight $\alpha_{ij} \propto \exp({e_j}^\mathrm{T} {\textbf{W}^l_k}^\mathrm{T} \textbf{W}_q e_i)$. $\textbf{W}_q$, $\textbf{W}^l_k$, and $\textbf{W}^l_v$ are the matrices that transform embedding vectors into ``query'', ``key'', and ``value'', respectively. After that, another GAT aggregates ${e'}_i^l$ from all groups into ${e'}_i$, which summarizes the local states of its neighbors and their relationships.

\subsection{Maximum Entropy Learning}

In maximum entropy learning, agents aim to learn a soft value function that combines each agent's total expected return and action entropy at each state. They train their policy $\pi$ with the object  $J(\pi)=\sum_{t=0}^{T}\mathbb{E}[r(t)+\alpha\mathcal{H}(\pi(\cdot|s(t)))]$, where $\mathcal{H}(\pi(\cdot|s(t)))$ is an entropy term and $\alpha$ is the temperature hyper-parameter that controls the stochasticity of $\pi$. Since they explore more widely, agents can avoid converging to non-optimal policies and improve their performance.

\section{Methods}\label{sec4}

In this section, we introduce our coordination control solution for large-scale multi-agent systems with continuous action spaces, including a graph learning network based on HGAT and a maximum entropy MARL approach named Entropy-enhanced Hierarchical graph Continuous Action Multi-Agent coordination control method (EHCAMA). 

\begin{figure}[H]
	\includegraphics[width=10.5 cm]{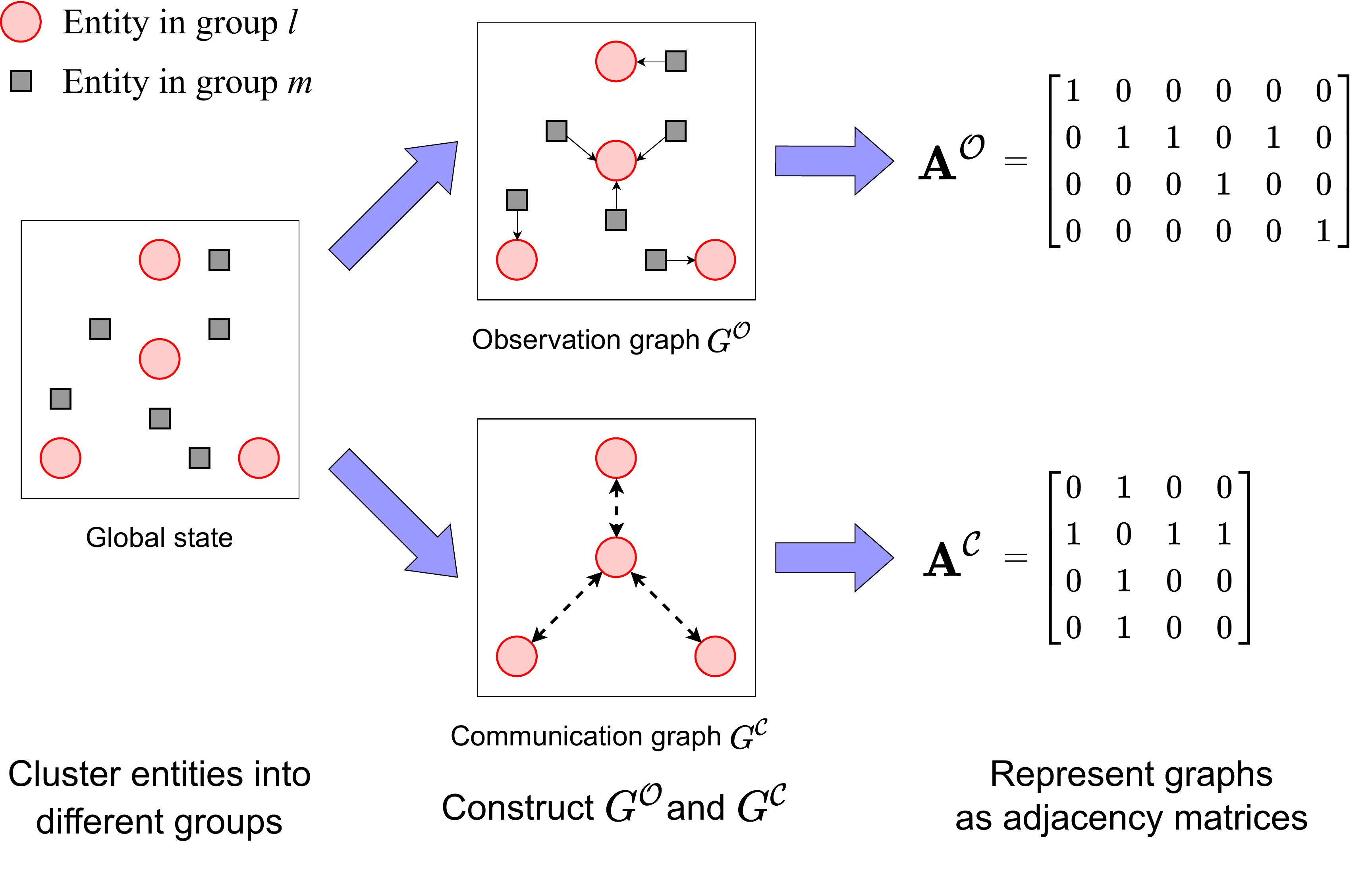}
	\caption{The graph representation of the global state in EHCAMA. \label{fig_group}}
\end{figure}

\begin{figure}[H]
	\includegraphics[width=10.5 cm]{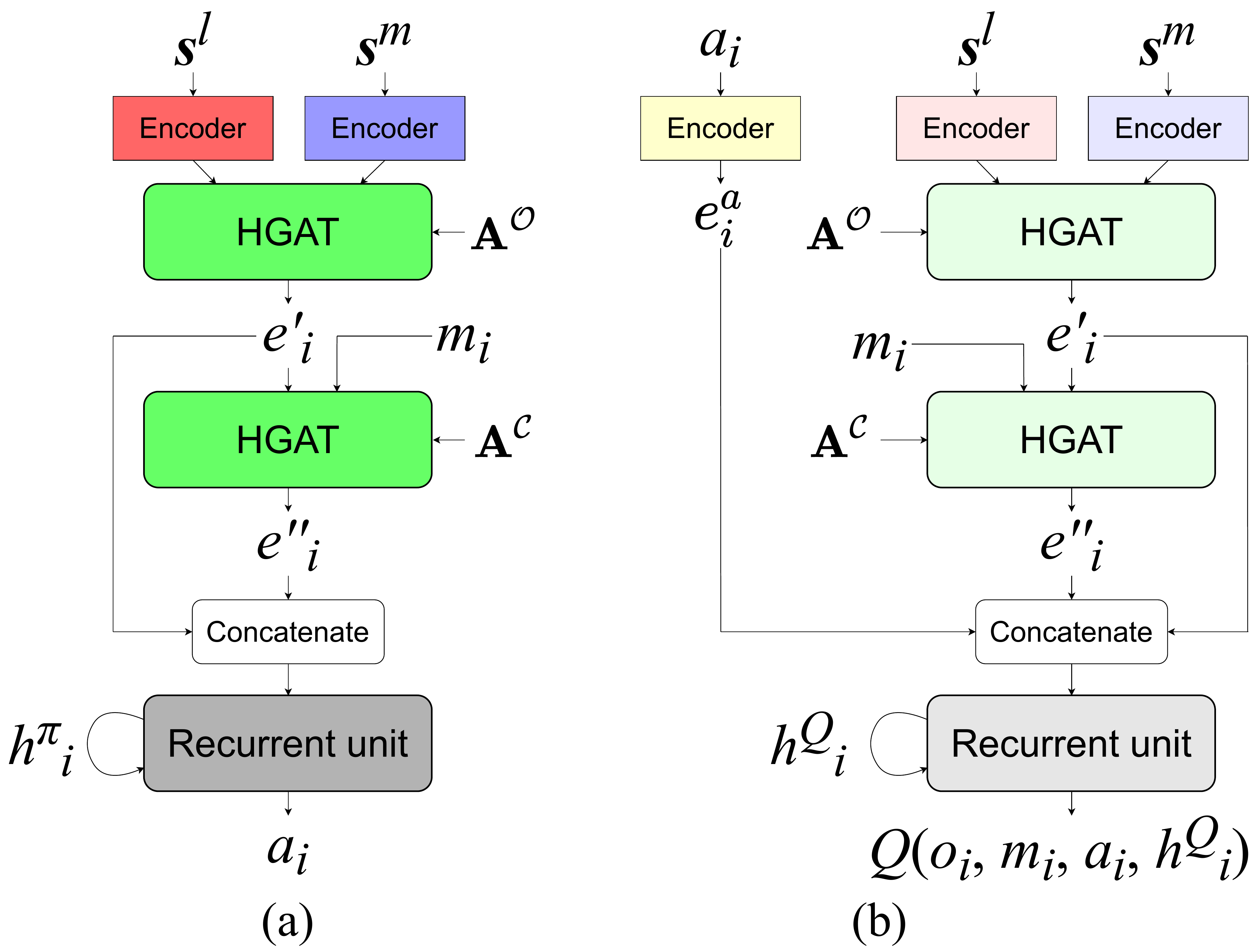}
	\caption{The network architecture of (a) the actor and (b) the critic in EHCAMA. \label{fig_network}}
\end{figure}

\subsection{Graph Representation and Network Architecture}

Figure~\ref{fig_group} shows the graph representation of the global state of the environment. The first step is to cluster $M$ entities (including agents, landmarks, etc.) in the environment into $K$ groups according to their features. Because of the high differences between these heterogeneous entities, it is more efficient for agents to treat them separately. The second step is to construct an observation graph $G^{\mathcal{O}}$ and a communication graph $G^{\mathcal{C}}$, where the nodes denote entities while the edges mean two of them are in their observation range and communication range, respectively. In the third step, EHCAMA transforms the global state into structured data, containing $\boldsymbol{s}$, $\textbf{A}^{\mathcal{O}}$, and $\textbf{A}^{\mathcal{C}}$, where $\boldsymbol{s}=(s_1, , \cdots, s_M)$ is a set of all local states, $\textbf{A}^{\mathcal{O}}$ and $\textbf{A}^{\mathcal{C}}$ are the adjacency matrices of $G^{\mathcal{O}}$ and $G^{\mathcal{C}}$.

The network architectures of the actor and critic in our proposed EHCAMA are shown in Figure~\ref{fig_network}. We share the parameters of the networks among the agents in the same group, thus improving the scalability of the multi-agent system. After each agent $i$ acquires its observation $o_i$ from the environment, the linear encoders in its actor network transform the raw data from different groups into embedding vectors with the same dimension. Then, the first HGAT layer uses $\textbf{A}^{\mathcal{O}}$ to select the vectors from agent $i$' observation range and processes them to summarize the information in $o_i$ into a high-dimensional embedding vector $e'_i$. Note that we employ multiple attention heads and replace the group-level GAT in the vanilla HGAT with a fully connected layer. 

In the inter-agent communication stage, each agent $i$ shares the information from $o_i$ by sending $e'_i$ to its neighbors, which helps it to cooperate better with other agents. Upon reception of message $m_i$, the second HGAT layer extracts features from $m_i$ and uses $\textbf{A}^{\mathcal{C}}$ to calculate $e''_i$ like the first layer. To accelerate converging, we concatenate $e'_i$ and $e''_i$ into a vector and input it into the recurrent unit.

\begin{figure}[H]
	\includegraphics[width=9 cm]{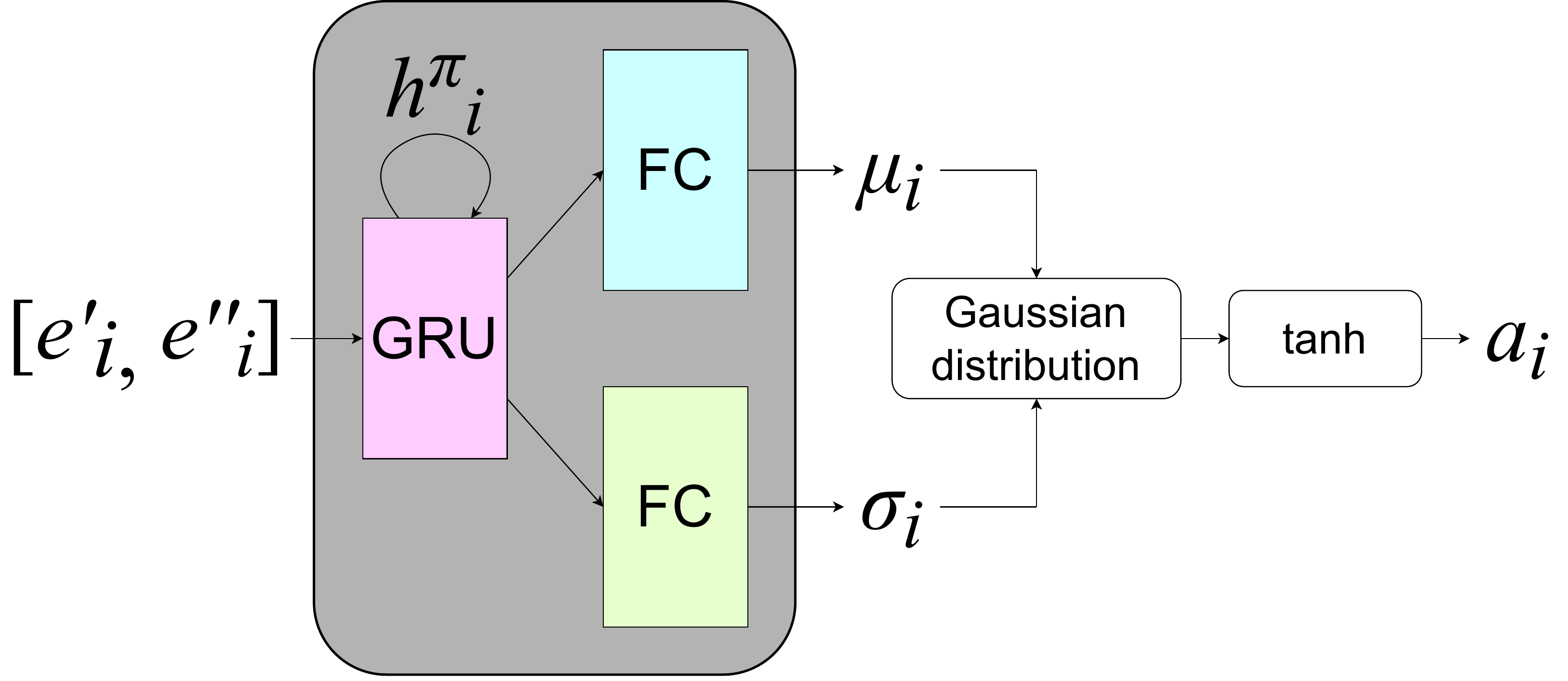}
	\caption{The structure of the recurrent unit in the stochastic actor. \label{fig_recurrent_unit}}
\end{figure}

In some prior actor-critic approaches for continuous action spaces, agents learn deterministic policies to determine their actions, such as DDPG. Unfortunately, since the deterministic actors only use an additional random noise for exploration, it is hard for them to stabilize \cite{henderson2018deep}. Thus, we introduce stochastic policies with Gaussian distributions to provide an intelligent exploration, thereby improving the performance of EHCAMA either in the training stage or the executing stage. As shown in Figure~\ref{fig_recurrent_unit}, we design a new recurrent unit that consists of a GRU layer and two fully connected layers. The GRU layer maintains the hidden states $h^{\pi}_i$ as the memory, which helps agent $i$ to recall the previous observations and record the new information from input vectors. After that, agent $i$ respectively calculates $\mu_i$ and $\sigma_i$ by two separate linear transforms and obtains the action distribution $\pi^l(o_i, m_i, h^{\pi}_i)$. In practice, we squash the Gaussian samples into $(-1, 1)$ as $a_i=\tanh(u_i)$, where $u_i \sim N(\mu_i, \sigma^2_i)$ \cite{haarnoja2018soft2}.

Similar to the actor network, the critic also uses HGAT layers to aggregate the information from observations and neighbors. Unlike our previous work \cite{chen2022scalable}, we cannot compute Q-values for each choice in continuous action spaces. Thus, the critic uses an individual linear encoder to process actions $a_i$, as shown in Figure~\ref{fig_network}(b). The critic's recurrent unit contains a GRU layer and a fully connected layer, which also maintains the hidden states. To reduce the overoptimism for Q-values, we use twin critic networks with different parameters and represent them as $Q^l_1$ and $Q^l_2$ for each group $l$ \cite{fujimoto2018addressing}. Their hidden states for each agent $i$ are denoted as $h^{Q_1}_i$ and $h^{Q_2}_i$, respectively.

\subsection{Training Process with Entropy-enhanced Optimization}

In standard actor-critic methods, the training strategy aims to maximize the total expected return $R$ of the agent. However, it cannot be used to train stochastic policies in EHCAMA because it does not consider the optimization of exploration. This fatal defect brings extreme instability to policies and causes performance deterioration in complex tasks. To this end, we introduce maximum entropy learning to optimize exploration. By maximizing $R_i$ and entropy $\mathcal{H}_i$ of each state, agent $i$ can adjust exploration to an appropriate degree. In addition, we apply target networks and a replay buffer to reinforce stability and sample-efficiency when training EHCAMA \cite{mnih2015human}.

At each timeslot, EHCAMA stores an experience $(\boldsymbol{s}, \textbf{A}^{\mathcal{O}}, \textbf{A}^{\mathcal{C}}, \boldsymbol{a}, \boldsymbol{r}, \boldsymbol{s}', {\textbf{A}'}^{\mathcal{O}}, {\textbf{A}'}^{\mathcal{C}}, \boldsymbol{h}, \boldsymbol{h}')$ into a replay buffer $B$ shared by $N$ agents, where $\boldsymbol{a}=(a_1, \cdots, a_N)$ and $\boldsymbol{r}=(r_1, \cdots, r_N)$. $\boldsymbol{s}'$, ${\textbf{A}'}^{\mathcal{O}}$ and ${\textbf{A}'}^{\mathcal{C}}$ represent the structured data of the next global state. $\boldsymbol{h}$ contains the hidden states of actors and critics in all agents and $\boldsymbol{h}'$ is the set of next hidden states. We set all hidden states to zero when we initialize a new episode.

In the training stage, we compute the objective for optimization by reusing the previous experiences sampled from $B$. To evaluate each global state in terms of $R_i$ and $\mathcal{H}_i$, we define a soft value function $V$ as:

\begin{equation}\label{eq:stochastic_value}
	\begin{aligned}
		V(o_i, m_i, a_i) =\min_{c=1,2}Q_c^l(o_i, m_i, a_i, h^{Q_c}_i|\theta^{Q_c}_l)- \alpha_l\log\pi^l(a_i|o_i, m_i, h^{\pi}_i, \theta^{\pi}_l)
	\end{aligned}
\end{equation}
where $\theta^{Q_c}_l$ and $\theta^{\pi}_l$ are the parameters of $Q_c^l$ and $\pi^l$, respectively. The temperature parameter $\alpha_l$ represents the importance of the entropy term for agent $i$ in group $l$.

For each critic network $Q_c^l$, we update parameters $\theta^{Q_c}_l$ by Bellman residual minimization as:

\begin{equation}\label{eq:loss_q_network}
	\begin{aligned}
		\mathcal{L}(\theta^{Q_c}_l)=\frac{1}{N_l}\sum_{i=1}^{N_l}\mathbb{E}[(y_i-Q_c^l(o_i, m_i, a_i, h^{Q_c}_i| \theta^{Q_c}_l))^2]
	\end{aligned}
\end{equation}
where $\mathcal{L}(\theta^{Q_c}_l)$ is the loss function and $N_l$ means the number of agents in group $l$. The target value $y_i$ is calculated as:

\begin{equation}\label{eq:target_value}
	\begin{aligned}
		y_i=r_i+\gamma V'(o'_i, m'_i, a'_i)
	\end{aligned}
\end{equation}
where $o'_i$ and $m'_i$  respectively denote $i$'s next observation and next received messages. $V'(o'_i, m'_i, a'_i)$ is the next soft value computed by the target networks ${Q'_c}^l$ and ${\pi'}^l$, whose parameters $\theta^{Q'_c}_l$ and $\theta^{\pi'}_l$ are transferred from $Q_c^l$ and ${\pi}^l$ via soft updates, respectively. The next action $a'_i$ is sampled from $ {\pi'}^l(o'_i, m'_i, {h'}^{\pi}_i)$.

The stochastic actor $\pi^l$ is optimized by the gradient:
\begin{equation}\label{eq:gradient_of_actor_network}
	\begin{aligned}
		\nabla_{\theta^{\pi}_l}\mathcal{J}(\theta^{\pi}_l)=\frac{1}{N_l}\sum_{i=1}^{N_l}\mathbb{E}[\nabla_{\theta^{\pi}_l}\pi^l(a_i|o_i, m_i, h^{\pi}_i, \theta^{\pi}_l)\nabla_{a_i}V(o_i, m_i, a_i)|_{a_i \sim \pi^l(o_i, m_i, h^{\pi}_i | \theta^{\pi}_l)}]
	\end{aligned}
\end{equation}

\section{Simulations}\label{sec5}

\subsection{The Experimental Environment}

\begin{figure}[H]
	\includegraphics[width=10.5 cm]{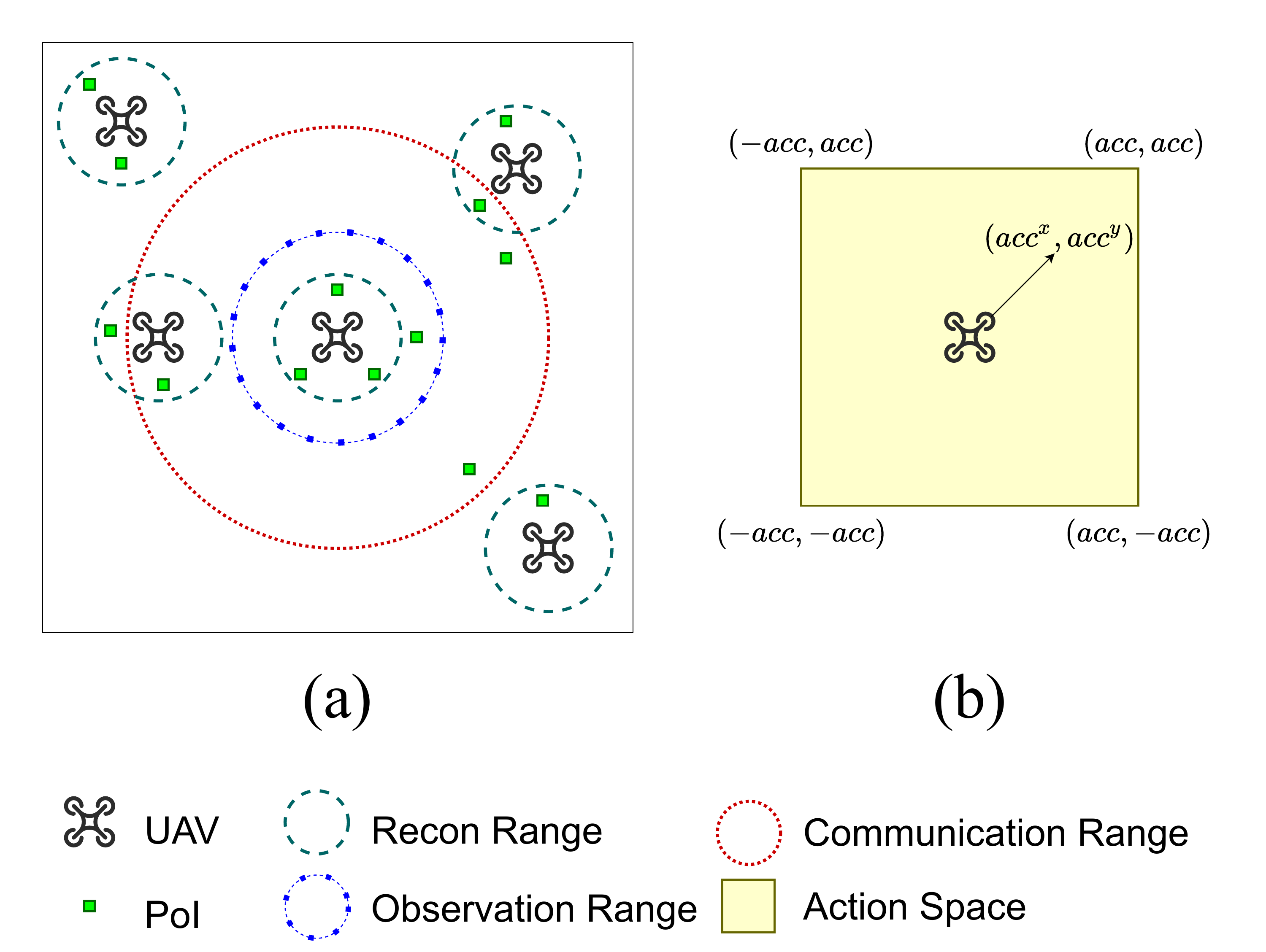}
	\caption{Illustrations of (a) the experimental environment and (b) the action space of UAVs. Note that the action space contain the border. \label{fig_scenario}}
\end{figure}

Based on our previous works \cite{ye2022multi} and \cite{chen2022scalable}, we construct a multi-agent environment named to test various capabilities of our proposed EHCAMA and baselines. As shown in Figure~\ref{fig_scenario}(a), this scenario consists of $n$ randomly distribute ground point-of-interest (PoIs) and $N$ reconnaissance UAVs at a fixed altitude. Each UAV $i$ takes off from a random position and scouts PoIs by covering them within its recon range. At each timeslot $t$, UAV $i$ obtains the local state (containing positions) of PoIs in its observation range and the messages from neighboring UAVs in its communication range. We assume that UAV $i$ sends its local state, consisting of its position and velocity, to its neighbors via inter-agent communication. Then, UAV $i$ utilizes received information to determine an acceleration $(acc^x_i, acc^y_i)$ as its action, where we replace finite choices in our previous works with a continuous action space shown in Figure~\ref{fig_scenario}(b). In the training stage, UAV $i$ is rewarded as \cite{chen2022scalable}: 

\begin{equation}\label{eq:reward_recon}
	\begin{aligned}
		r_i=\dfrac{\eta_1\times r_{indv}+\eta_2\times r_{shared}}{E_i(t)}-p_i \ , \quad{\rm where}\quad
		r_{indv}&=\left\lbrace
		\begin{array}{lr}
			-1\ , & \quad if \ n_{indv}=0\\
			n_{indv}\ , & \quad otherwise
		\end{array}
		\right. \ ,\\
		r_{shared}&=\left\lbrace
		\begin{array}{lr}
			\quad 0 \ , & if \ n_{shared}=0\\
			\dfrac{n_{shared}}{N_{share}}, & otherwise
		\end{array}
		\right. , \\
		E_i(t)&=E_h+\dfrac{v_i(t)}{v_{max}}E_m
	\end{aligned}
\end{equation}
where $n_{indv}$ means the number of PoIs covered by UAV $i$ individually and $n_{shared}$ denotes the number of PoIs overlapped with $N_{share}$ neighboring UAVs. $E_i(t)$ represents the energy consumption of $i$ at $t$, where $E_h$ and $E_m$ are the energy consumed for hovering and moving, respectively. $v_i(t)$ is $i$'s velocity at $t$ while $v_{max}$ is the maximum speed of UAVs.  The penalty factor $p_i=1$ when $i$ flies outside the target area, otherwise $p_i=0$. We set the hyper-parameter $\eta_1$ to 1 and $\eta_2$ to 0.1.

We deploy an MARL method to coordinate UAVs to execute cooperative reconnaissance tasks, whose quality is evaluated based on coverage $C$, fairness $F$, and energy consumption $E$. We use an overall indicator $CFE$ that combines the score of $C$, $F$, and $E$ as the major metric in evaluation, which is defined as:

\begin{equation}\label{eq:overall_indicator}
	\begin{aligned}
		CFE &= \dfrac{C\times F}{E}, \quad{\rm where}\
		C = \frac{1}{T}\sum_{t=1}^{T} \dfrac{n_C(t)}{n},\
		F = \dfrac{(\sum_{j=1}^{n}c_j)^2}{n\sum_{j=1}^{n}c_j^2},\
		E = \frac{1}{T\times N}\sum_{t=1}^{T} \sum_{i=1}^{N}E_i(t)
	\end{aligned}
\end{equation}
where we calculate $F$ based on Jain’s fairness index \cite{jain1984quantitative}. $T$ means the total timeslots of a task, $n_C(t)$ denotes the number of PoIs covered by UAVs at timeslot $t$, and $c_j$ is the coverage time of PoI $j$ at the end of a task.

In simulations, we set the target area as $200 \times 200$ units and the number of PoIs as 120. $v_{max}$ is 10 units while $E_h$ and $E_m$ are both 0.5. The recon range, observation range, and communication range are set to 10, 15, and 30 units, respectively. In addition, we set $acc$ in Figure~\ref{fig_scenario}(b) as 4 units and the length of each episode is 100 timeslots. 

\subsection{Experimental Setup and Baseline Methods}

We implement our proposed EHCAMA with Pytorch and simulate its performance on an Ubuntu 18.04 server with 2 NVIDIA RTX 3080 GPUs. Empirically, we set the hyper-parameters of EHCAMA as follows. We apply Adam as the optimizer and set learning rate to 0.001. The discount factor $\gamma$ is 0.95 and the soft update rate $\tau$ is 0.01. The number of units in each fully connected layer and GRU layer is 256 while each HGAT layer contains 4 attention heads. We set the capacity of the replay buffer as 50K and the size of a minibatch as 128. The parameters of networks are updated 4 times every 100 timeslots when training.

As the experimental baselines, we consider four MARL methods, DDPG, SAC, HAMA, and MADDPG. Moreover, we emulate DDPG to implement a deterministic variant of our method named ``DHCAMA'' for comparing. We summarize the comparison of our method and baselines in Table~\ref{tab_comparison}. In non-HGAT-based approaches, the observation of each agent $i$ is a high-dimensional vector that contains all local states of $i$'s observed entities. Each method is trained for 100K episodes and tested for 10K episodes. We set the temperature hyper-parameter $\alpha$ for EHCAMA and SAC to 2.5 in simulations.

\begin{table}[H] 
	\setlength{\tabcolsep}{4mm}\caption{Comparison of various MARL methods. \label{tab_comparison}}
	\begin{threeparttable}
		\begin{tabular}{ccccccc}
			\toprule
			\textbf{Methods}		&
			\textbf{Action Policy}	&
			\textbf{E}\tnote{1}		&
			\textbf{H}\tnote{2}		&
			\textbf{S}\tnote{3}		&
			\textbf{R}\tnote{4}		&
			\textbf{C}\tnote{5}\\
			
			\midrule
			
			EHCAMA (ours) & Stochastic & \checkmark & \checkmark & \checkmark & \checkmark & \checkmark \\
			DHCAMA (ours) & Deterministic &  & \checkmark & \checkmark & \checkmark & \checkmark \\
			DDPG & Deterministic &  &  & \checkmark &  &  \\
			SAC & Stochastic & \checkmark &  & \checkmark &  &  \\
			HAMA & Deterministic &  & \checkmark & \checkmark  &  &  \\
			MADDPG & Deterministic &  &  &  &  &  \\
			
			\bottomrule
		\end{tabular}
		
		\begin{tablenotes}
			\footnotesize
			\item[1] Maximum Entropy Learning
			\item[2] HGAT
			\item[3] Parameter Sharing
			\item[4] Recurrent Unit
			\item[5] Inter-agent Communication
		\end{tablenotes}

	\end{threeparttable}
\end{table}

\subsection{Experimental results}

\begin{figure}[H]
	\begin{tabular}{c}
		\includegraphics[width=7. cm]{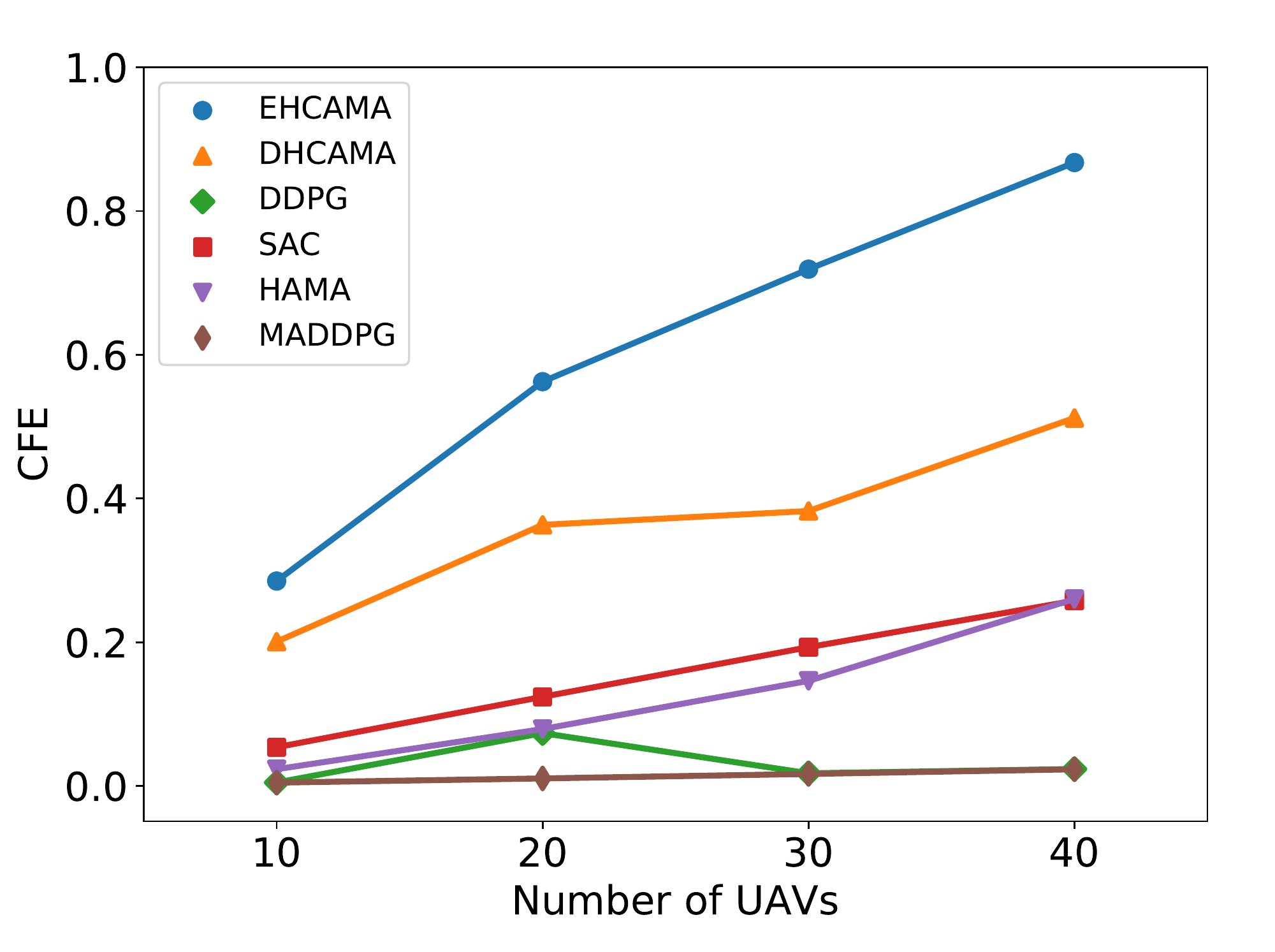}
		\includegraphics[width=7. cm]{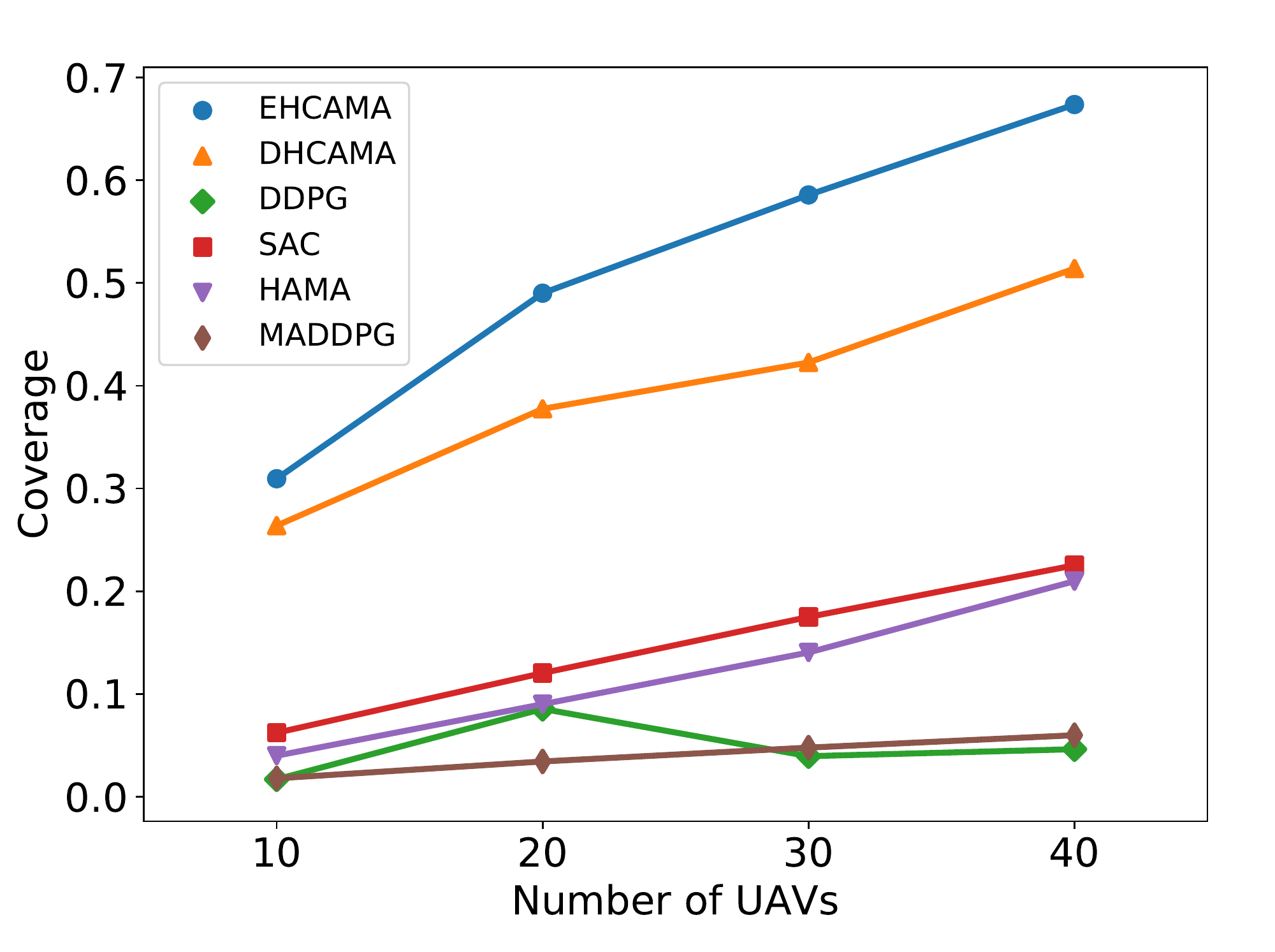}\\
		\includegraphics[width=7.cm]{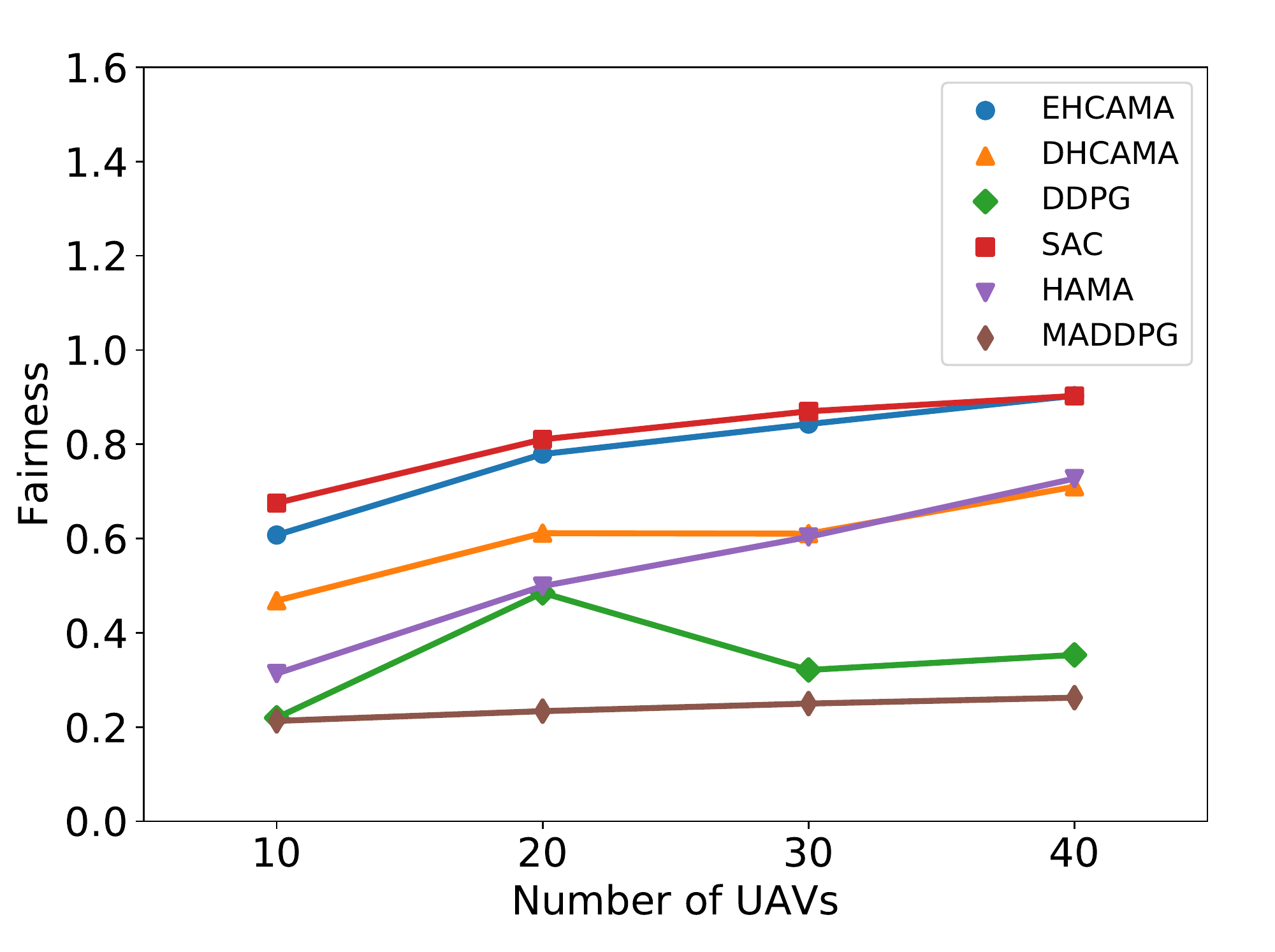}
		\includegraphics[width=7.cm]{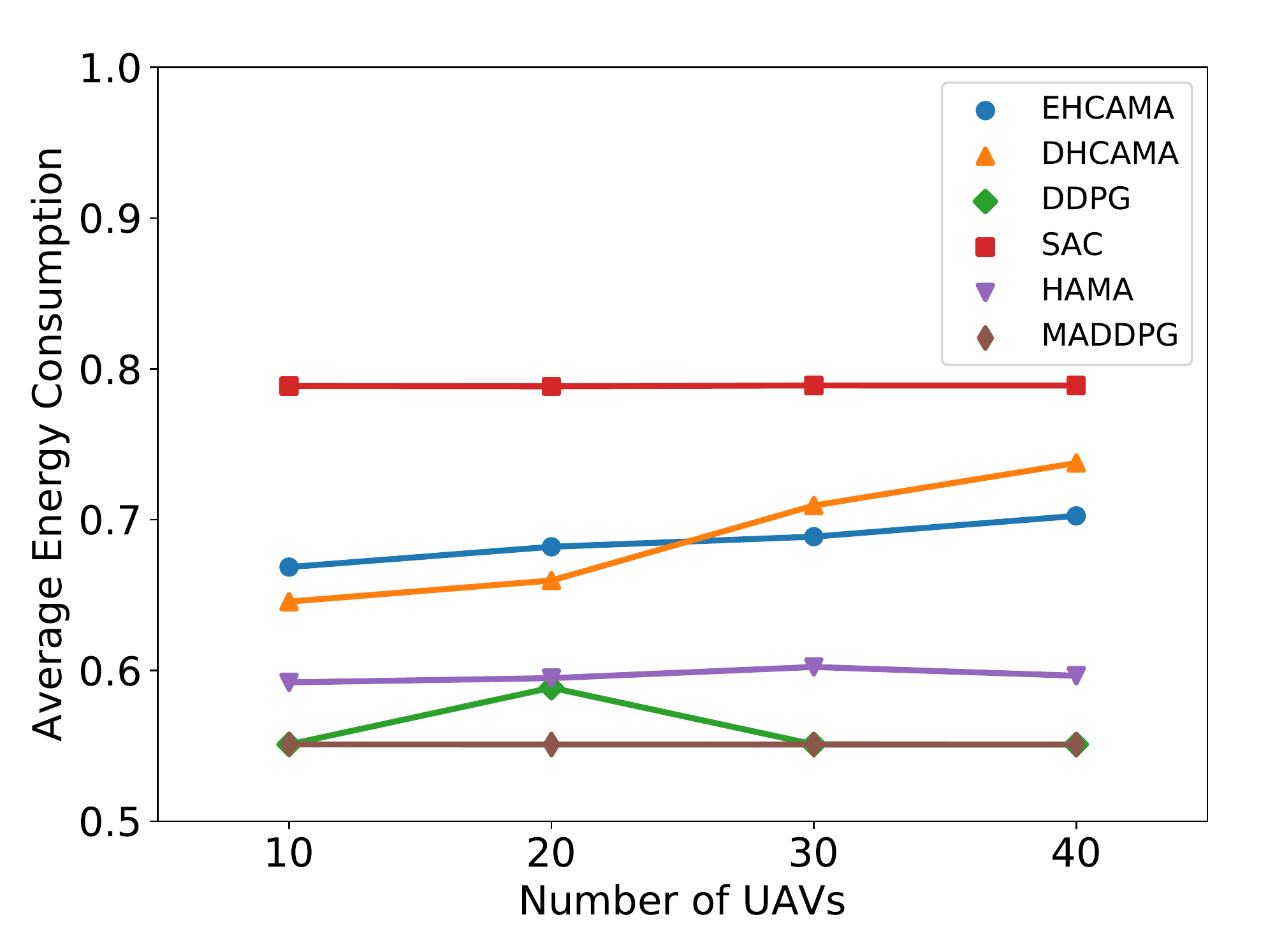}
	\end{tabular}
	\caption{Impact of the number of UAVs on four metrics of all methods.}
	\label{fig_result_uav_recon_1}
\end{figure}

We test the scalability of each method and show the impact of the number of UAVs on their four metrics (including $CFE$, coverage, fairness, and energy consumption) in 	Figure~\ref{fig_result_uav_recon_1}. As observed in Figure~\ref{fig_result_uav_recon_1}(a), EHCAMA's performance in terms of $CFE$ is consistently superior to others in various scale tasks, while DHCAMA is the second best. We consider the reasons why EHCAMA succeeds as follows.

First, because HGAT can model the hierarchical relationship in observation, UAVs can extract features from structured data represented as graphs, which is better than raw vectors in non-HGAT-based approaches. Second, the inter-agent communication and the hidden states in recurrent units provide extra information from their neighbors and memories to UAVs, which improves the performance of UAVs in cooperation. Finally, maximum entropy learning makes the stochastic policies tractable for exploring. Intelligent exploration helps UAVs to learn the reward distribution on continuous action space and prevents them from optimizing their policies to local optimums, thereby bringing stability and robustness when executing and training.

As a result, in Figure~\ref{fig_result_uav_recon_1}(b), both implements of our method obtain higher coverage than baselines. By efficiently learning the distribution of PoIs from their observations, neighbors, and memories, UAVs can cooperate with their companions to cover more PoIs. Further, EHCAMA performs even better than the deterministic variant. We consider this as the contribution of the entropy enhanced optimization. For fairness and energy consumption, we observe two extreme cases in Figure~\ref{fig_result_uav_recon_1}(b) and (c). Compared with EHCAMA, UAVs in deterministic baselines prefer to save energy instead of improving coverage and fairness. We hypothesize that these approaches output locally optimal policies where UAVs maximize their total expected return by reducing energy consumption. On the contrary, UAVs in SAC consume more energy than EHCAMA to cover PoIs fairly. Although maximum entropy learning encourages exploration, they cannot optimize their flight trajectory by learning from raw observation vectors, so they cover each PoI for a nearly equal but short time. 

\begin{figure}[H]
	\begin{tabular}{c}
		\includegraphics[width=7. cm]{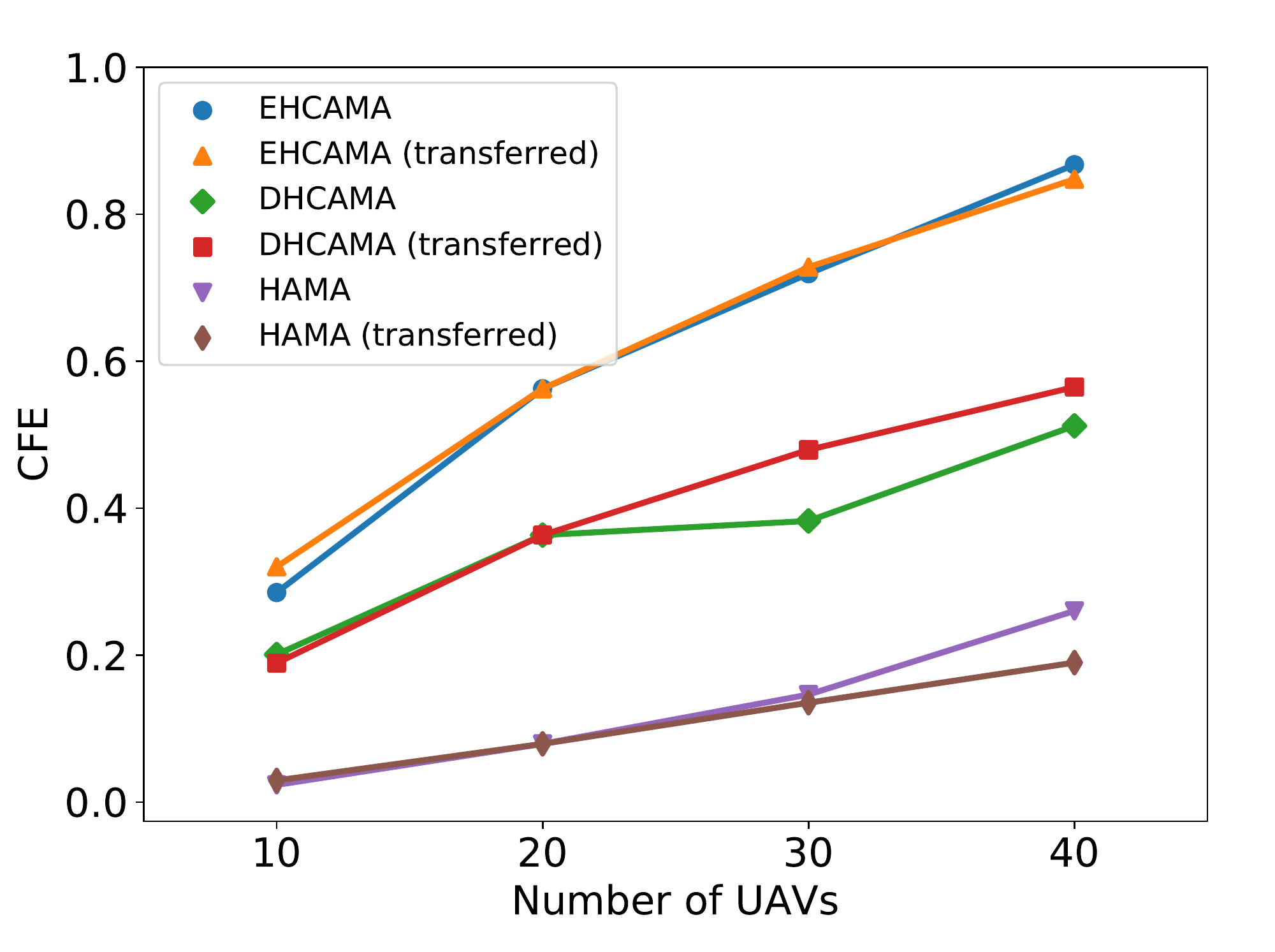}
		\includegraphics[width=7. cm]{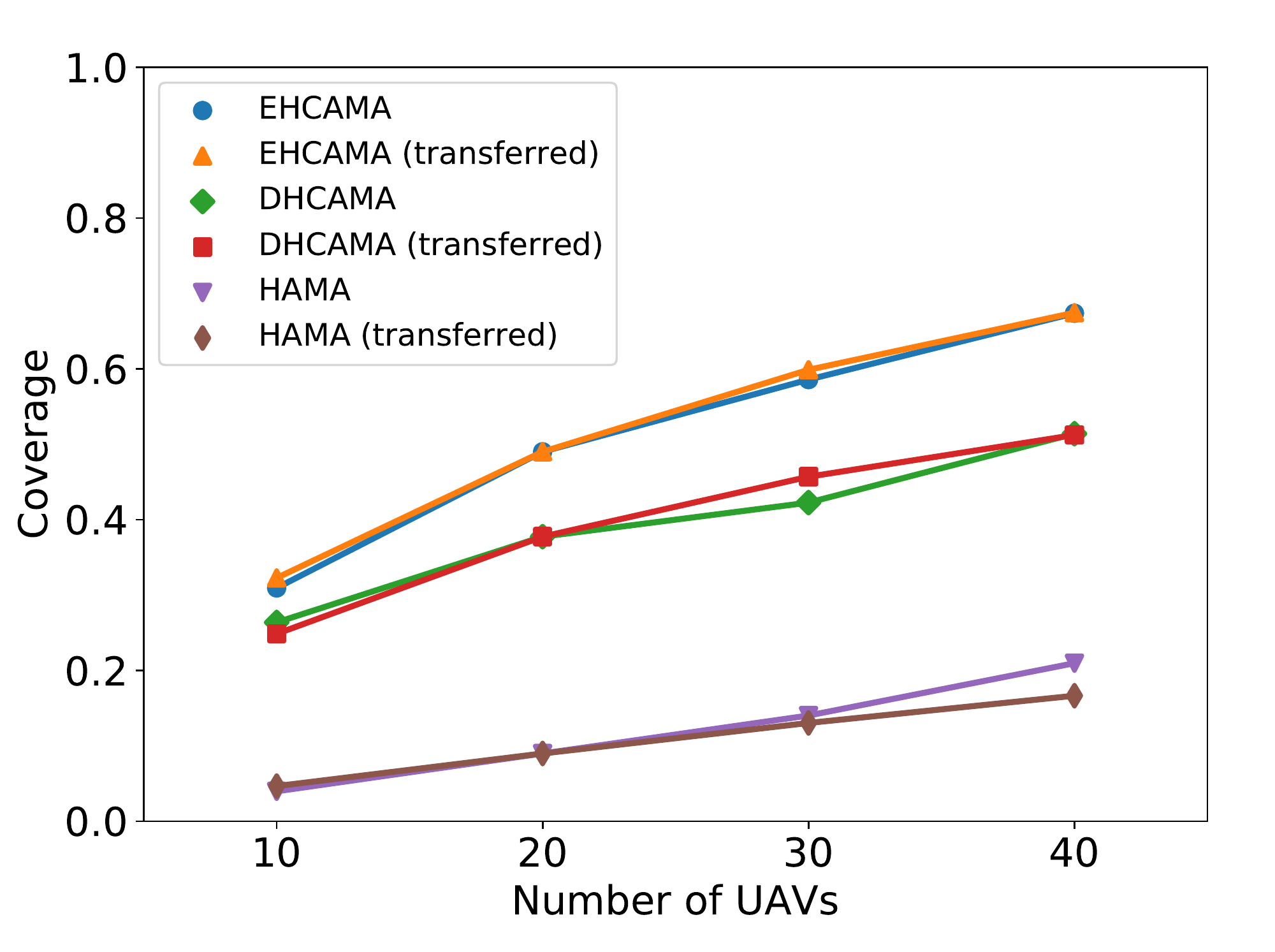}\\
		\includegraphics[width=7.cm]{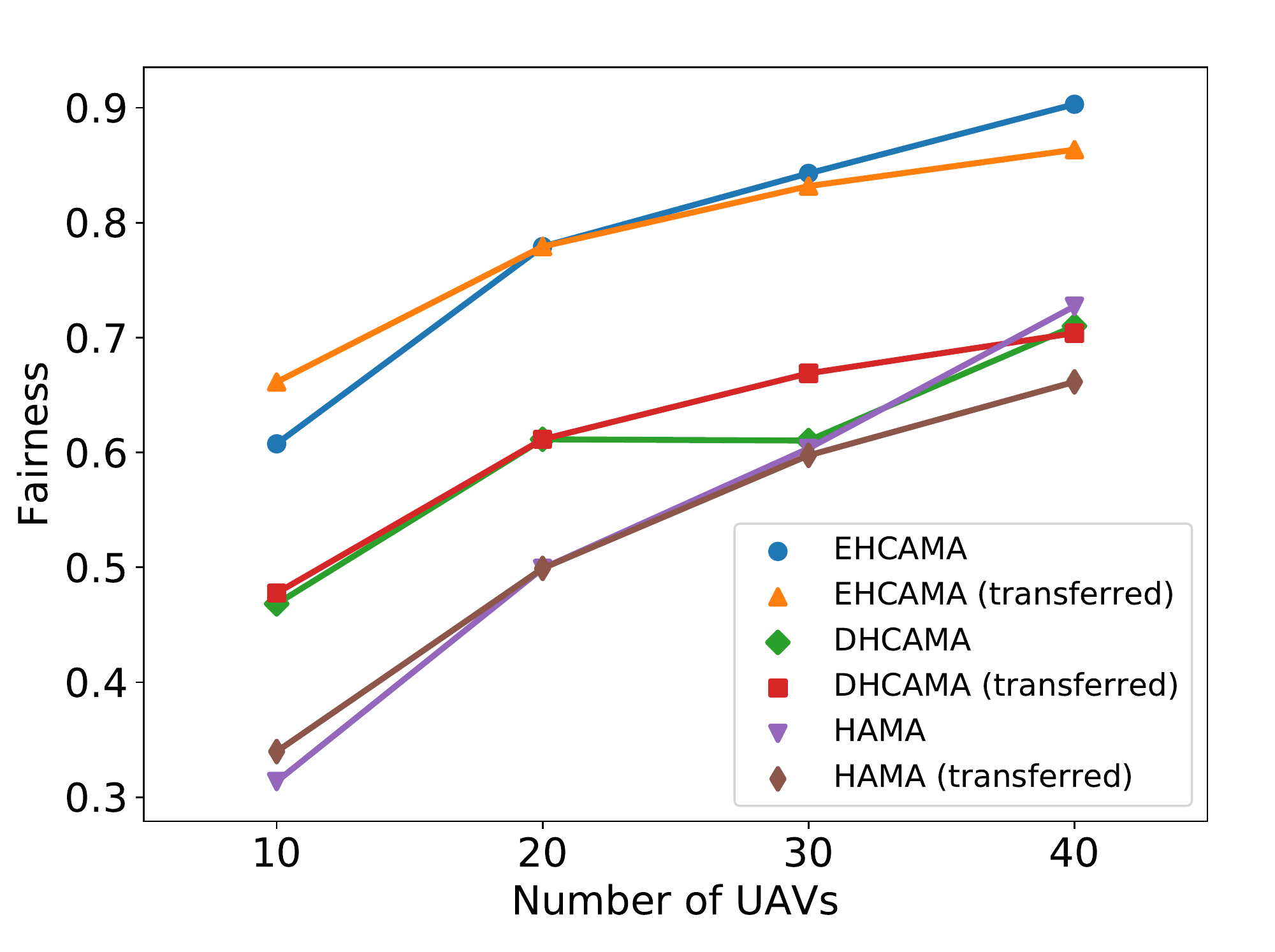}
		\includegraphics[width=7. cm]{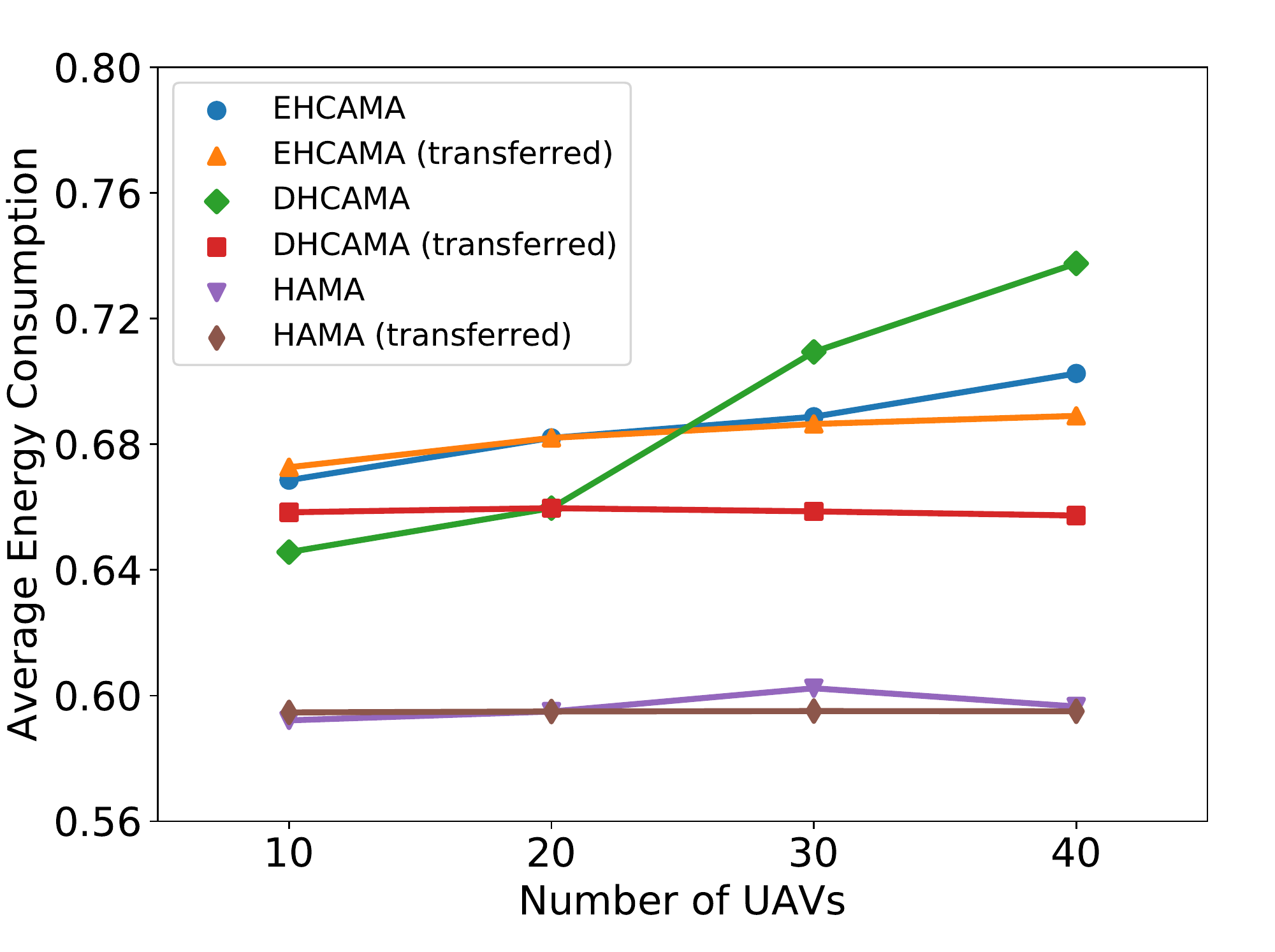}
	\end{tabular}
	\caption{Impact of the number of UAVs on four metrics of transfer learning.}
	\label{fig_result_uav_recon_2}
\end{figure}

In addition, MADDPG does not share the network parameters among UAVs. Therefore, it has to optimize an individual policy for each UAV, which spends a lot of time on training but performs worse than our method in terms of scalability.

We test the transferability of three HGAT-based methods and show the experimental results in Figure~\ref{fig_result_uav_recon_2}, where we train transferred policies with 20 UAVs. The performance of EHCAMA does not appear to degradation when transferring to different scale tasks. Moreover, it outperforms deterministic methods in terms of $CFE$, coverage, and fairness, whether they transfer or not.

\subsection{Ablation Study}

To verify the benefit of stochastic policies and maximum entropy learning, we compare the performance of EHCAMA with DHCAMA and two stochastic variants described as follows:

\begin{itemize}
	\item Variant A: disabling the entropy term in Equation~\ref{eq:stochastic_value} (i.e. $\alpha_l=0$);
	\item Variant B: optimizing the standard maximization objective with on-policy learning based on advantage actor-critic (A2C) \cite{degris2012model}.
\end{itemize}

\begin{figure}[H]
	\begin{tabular}{c}
		\includegraphics[width=9 cm]{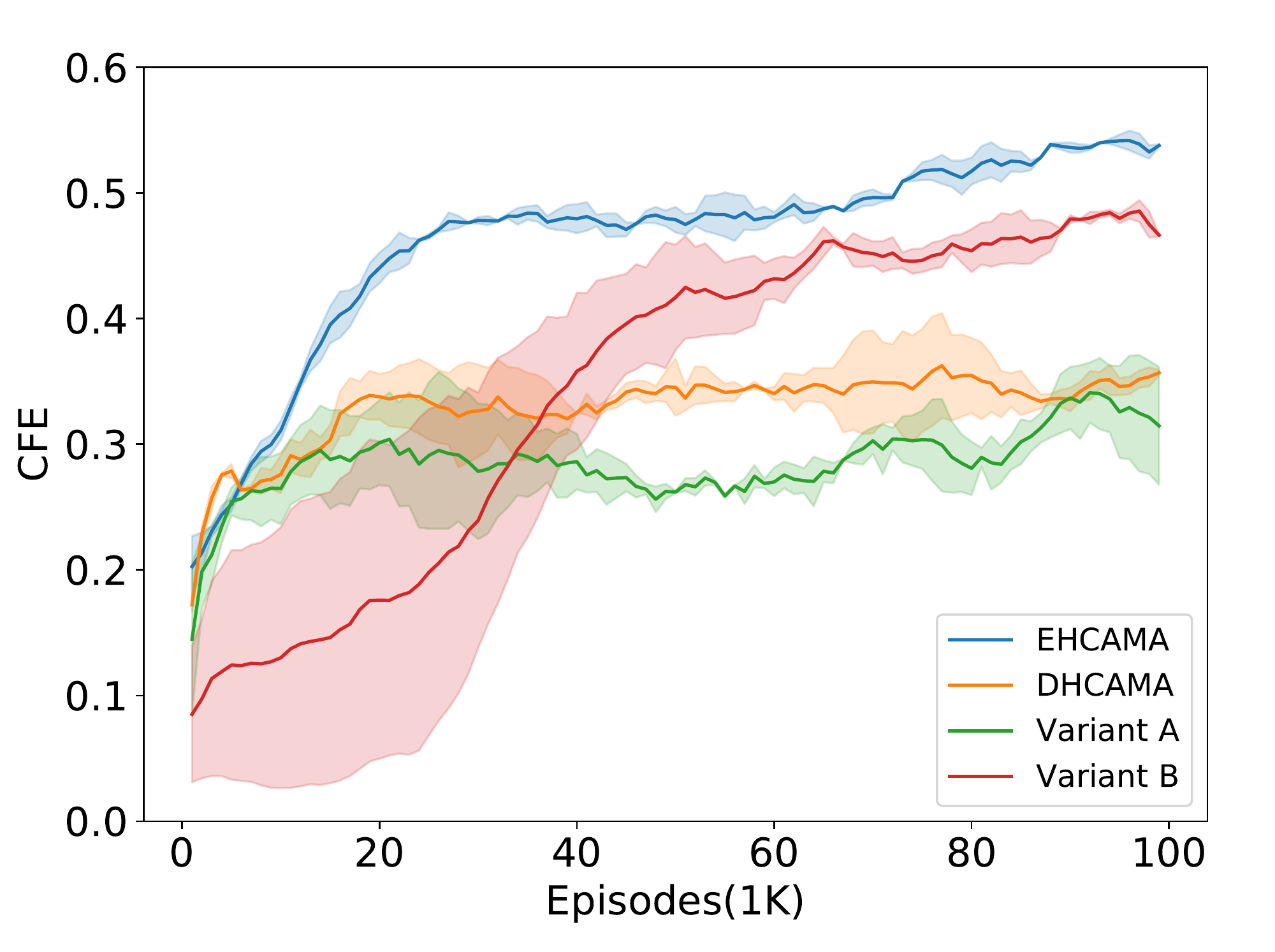}
	\end{tabular}
	\caption{Learning curve of each variant (20 UAVs, 3 different random seeds).}
	\label{fig_result_learning_curve_1}
\end{figure}

We train each variant with three different random seeds for 100K episodes and show their learning curve in Figure~\ref{fig_result_learning_curve_1}. From the experimental results, we observe that EHCAMA outperforms all variants in $CFE$ because it applies maximum entropy learning and a replay buffer to train policies. DHCAMA and Variant A optimize the standard maximum expected return objective in the training stage, which brings instability to their policies. Although Variant A employs stochastic control, it still obtains worse $CFE$ compared with EHCAMA. Variant B performs better than DHCAMA since it learns more stable policies by on-policy learning. However, due to its poor sample efficiency, it needs more samples to converge in large-scale multi-agent tasks and shows its brittleness to the random seed.

\section{Conclusion}\label{sec6}

In this paper, we propose an entropy-enhanced MARL method named EHCAMA to solve the control problem in large-scale multi-agent systems with continuous action spaces. By employing entropy-enhanced optimization based on maximum entropy learning, agents in EHCAMA can learn stable stochastic policies to explore the environment intelligently. From experimental results, we observe that the performance of EHCAMA in terms of CFE is superior to all baselines (including DDPG, SAC, HAMA, and MADDPG) as well as the deterministic variant DHCAMA. Meanwhile, our method also shows scalability and transferability in various scale multi-agent systems. As a future work, we will investigate more intelligent exploration in MARL and try to improve the training strategy with adaptive temperature parameter.

\bibliographystyle{plainnat}
\bibliography{bibsample}
\end{document}